\documentstyle[12pt,lathuile]{article}
\begin{document}
\title{  SUSY DARK MATTER}
\author{
Achille Corsetti and Pran Nath     \\
{\em Department of Physics, Northeastern University, } \\
{\em Boston, MA 02115-5005, USA}              
}

\maketitle
\baselineskip=14.5pt
\begin{abstract}
Several new effects have been investigated in recent analyses of 
supersymmetric dark matter. These include the effects of the 
uncertainties of wimp velocity distributions, of the uncertainties 
of quark densities, of large CP violating phases, of nonuniversalities 
of the soft SUSY breaking parameters at the unification scale and 
of coannihilation on supersymmetric dark matter. We review here 
some of these with emphasis on the effects of nonuniversalities of the 
gaugino masses at the unification scale on the neutralino-proton 
cross-section from scalar interactions. The review encompasses several 
models where gaugino mass nonuniversalities occur including SUGRA models 
and D brane models. One finds that gaugino mass nonuniversalities can 
increase the scalar cross-sections by as much as a factor of 10 and 
also significantly extend the allowed range of the neutralino mass 
consistent with constraints up to about 500 GeV. These results have 
important implications for the search for supersymmetric dark matter.

\end{abstract}
\baselineskip=17pt
\newpage
\section{INTRODUCTION}
Over the recent past there has been considerable experimental 
activity in the direct detection of dark matter\cite{dama,cdms} 
and further progress is expected in the 
ongoing experiments\cite{dama,cdms,spooner}  
and new experiments that may come online in the future\cite{baudis}.
At the same time there have been several theoretical developments
which have shed light on the ambiguities and possible corrections 
that might be associated with the predictions on supersymmetric
dark matter.  These consist of the effects on the dark 
matter analyses of wimp velocity\cite{bottino1,roskowski,corsetti}
and of the rotation of the galaxy\cite{kk},
the effects of the uncertainties of 
quark densities\cite{bottino2,efo,corsetti2} and  the uncertainties 
of the SUSY parameters\cite{chung}, 
effects of large CP violating phases\cite{fos,cin},
effects of scalar nonuniversalities\cite{nonuni}, 
 effects of nonuniversalities of gaugino
masses\cite{corsetti2} and  effects of 
coannihilation\cite{coanni}. In this
paper we will discuss some of these briefly but mainly focus on the 
effects of nonuniversalities
of the gaugino masses on dark matter. 
In the Minimal Supersymmetric Standard Model (MSSM) 
there are 32 supersymmetric particles and with 
R parity conservation the lowest mass supersymmetric particle (LSP)
is absolutely stable. In many unified models, such as in the
 SUGRA models\cite{chams}, 
 one finds that the lightest neutralino is the LSP over  most
of the parameter space of the model. Thus the lightest neutralino
is a candidate for cold dark matter. The quantity that constrains
supersymmetric models is $\Omega_{\chi} h^2$ 
 where  $\Omega_{\chi}=\rho_{\chi}/\rho_c$ where $\rho_{\chi}$ is the
density of relic neutralinos at the current temperatures,
and  $\rho_c=3H_0^2/8\pi G_N$ is the critical
matter density, and  h is the Hubble parameter $H_0$
in units of 100 km/sMpc. The most recent measurements of h from the
Hubble  Space Telescope give\cite{freedman}
\begin{equation}
 h=0.71\pm 0.03\pm 0.07
\end{equation}
Similarly the most recent analyses of $\Omega_m$ give\cite{lineweaver} 
\begin{equation}
\Omega_m=0.3\pm 0.08
\end{equation}
If we assume that the component of $\Omega_B$ in $\Omega_m$ is 
 $\Omega_B\simeq 0.05$ which appears reasonable, then this leads
 to the result  $\Omega_{\chi} h^2=0.126\pm 0.043$. Perhaps 
 a more cautious choice of the range would be a $\sim 2 \sigma$
 range which gives
 \begin{equation}
  0.02\leq \Omega_{\chi} h^2\leq 0.3
 \end{equation}
 The quantity of interest theoretically is 
\begin{eqnarray}
\Omega_{\chi} h^2\cong 2.48\times 10^{-11}{\biggl (
{{T_{\chi}}\over {T_{\gamma}}}\biggr )^3} {\biggl ( {T_{\gamma}\over
2.73} \biggr)^3} {N_f^{1/2}\over J ( x_f )}
\end{eqnarray}
Here $T_f$ is the freeze-out temperature, $x_f= kT_f/m_{\tilde{\chi}}$
where  k is the Boltzman constant, $N_f$ is the number of 
degrees of freedom at the time of the  freeze-out,
 $({{T_{\chi}}\over {T_{\gamma}}})^3$ is the
reheating factor, and  $J(x_f)$ is given by
\begin{equation}
J~ (x_f) = \int^{x_f}_0 dx ~ \langle~ \sigma \upsilon~ \rangle ~ (x) GeV^{-2}
\end{equation} 
where $<\sigma v>$ is the thermal average with $\sigma$ the 
neutralino annihilation cross-section and v the neutralino relative
velocity. 

 \section{Detection of Milky Way wimps}
 Both direct and indirect methods are desirable and complementary 
 for the detection of Milky Way wimps.  We shall focus here
 on the direct detection. In this case the fundamental
 detector is the quark and the relevant interactions are
 the supergravity neutralino-quark-squark interactions. 
 The scattering of neutralinos from quarks contains squark poles 
 in the s channel and the Z boson and the Higgs boson ($h,H^0,A^0$) poles
 in the t channel. Since the wimp scattering from quarks is occuring 
 at rather low energies one may, to a good approximation, integrate
 on the intermediate squark, Z and Higgs poles to obtain a low  
 energy effective  Lagrangian which gives a four-Fermi interaction
 of the following form
\begin{equation}
{\cal L}_{eff}=\bar{\chi}\gamma_{\mu} \gamma_5 \chi \bar{q}
\gamma^{\mu} (A P_L +B P_R) q+ C\bar{\chi}\chi  m_q \bar{q} q
+D  \bar{\chi}\gamma_5\chi  m_q \bar{q}\gamma_5 q
\end{equation}
 The contribution of D is generally small and
 thus the scattering is effectively governed by the terms A,B
 and C. Analysis of dark matter is affected by several factors.
 We discuss these briefly below.
 
 \subsection{Uncertainties in wimp density and velocity}
Two of the quantities that  control the detection of dark 
matter are the wimp mass density  and the wimp velocity.
Estimates of Milky Way wimp density lie in the  
 range\cite{gates}  $\rho_{\chi} =  (0.2-0.7) GeVcm^{-3}$
  and the event rates in the direct detection depend
 directly on this density.  A second important factor regarding wimps
 that enters in the dark matter analyses is the wimp velocity.
 One typically assumes  a Maxwellian velocity 
distribution for the wimps and the current estimates for the rms wimp 
velocity give $v=270 ~km/s$ with, however, a significant 
uncertainty. Estimates for the uncertainty lie in the range of
$\pm 24$ km/s to $\pm 70$ km/s \cite{knapp}. A reasonable estimate 
then is that the rms wimp velocity lies in the range\cite{knapp} 
\begin{equation}
 v=270\pm 50 ~km/s
 \end{equation}
 Analyses including the wimp velocity variations show that the
 detection rates can have a significant variation, i.e.,
  a factor of 2-3 on either side of the central values\cite{corsetti}.
\subsection{Effects of uncertainties of  quark densities}
The scattering of neutralinos from quarks are dominated by
the scalar interaction which is controlled by the term C in Eq.(6). 
The dominant part of the scattering thus arises from the scalar
part of the $\chi -p$  cross-section  which is given by
 \begin{equation}
 \sigma_{\chi p}(scalar)=\frac{4\mu_r^2}{\pi}
 (\sum_{i=u,d,s}f_i^pC_i+\frac{2}{27}(1-\sum_{i=u,d,s}f_i^p)
 \sum_{a=c,b,t}C_a)^2
 \end{equation}
 Here $\mu_r$ is the reduced mass in the $\chi -p$ system
 and $f_i^p$ (i=u,d,s quarks) are quark densities inside the 
 proton defined by
 \begin{equation}
 m_pf_i^p=<p|m_{qi}\bar q_iq_i|p>
 \end{equation}
There are significant uncertainties in the determination of $f_i^p$.
To see the range of these uncertainties it is useful to 
parametrize the  quark densities so that\cite{corsetti2}
\begin{eqnarray} 
f_u^p=\frac{m_u}{m_u+m_d}(1+\xi)\frac{\sigma_{\pi N}}{m_p}\nonumber\\
f_d^p=\frac{m_d}{m_u+m_d}(1-\xi)\frac{\sigma_{\pi N}}{m_p}\nonumber\\
f_s^p=\frac{m_s}{m_u+m_d}(1-x)\frac{\sigma_{\pi N}}{m_p}
\end{eqnarray}
where we have defined
 \begin{eqnarray}
 x=\frac{\sigma_0}{\sigma_{\pi N}}=
 \frac{<p|\bar uu+\bar dd-2\bar ss|p>}{<p|\bar uu+\bar  dd|p>},~~ 
 \xi=
 \frac{<p|\bar uu-\bar dd|p>}{<p|\bar uu+\bar  dd|p>}\nonumber\\
 \sigma_{\pi N}= <p|2^{-1}(m_u+m_d)(\bar uu+\bar dd|p>
\end{eqnarray}
The current range of determinations of 
$\sigma_{\pi N}$ give\cite{corsetti2}
\begin{equation}
 \sigma_{\pi N}= 48\pm 9~ MeV, ~x=0.74\pm 0.25, ~\xi=0.132\pm 0.035
\end{equation}
With the above range of errors one finds that $f_i^p$ lie in the 
range 
\begin{equation}
f_u^p=0.021\pm 0.004,~~
f_d^p=0.029\pm 0.006,~~
f_s^p=0.21\pm 0.12
\end{equation}
Of these the errors in $f_s^p$ generates the largest variations.
A detailed analysis shows that the  scalar  cross-section can vary 
by a factor of 5 in either direction due to errors in the quark 
densities\cite{corsetti2}.

\subsection{CP violation effects on dark matter}
The soft SUSY breaking parameters that arise  in supersymmetric
theories after spontaneous breaking are in general complex 
with phases O(1) and can lead to large electric dipole 
moment of the electron and of the neutron in conflict with 
current experiment. Recently a cancellation
mechanism was proposed as a possible solution to this 
problem\cite{in1,in2}.
 With the cancellation mechanism the
total EDM of the electron and of the neutron can be in conformity 
with data even with phases O(1) and  sparticle masses which are
relatively light. The presence of large CP phases can affect dark 
matter and other low energy phenomena. 
Effects of CP violating phases on dark matter have been investigated 
for some time\cite{fos} and more recently such analyses have been extended 
to determine the effects of large CP phases under the cancellation 
mechanism\cite{cin}. 
One finds that the EDM constraints play a crucial role in these
analyses. Thus in the absence of CP violating phases one finds 
that the $\chi -p$ cross-section can change by orders of magnitude
when plotted as a function of $\theta_{\mu}$ (the phase of $\mu$).
 These effects are, however, significantly reduced when the 
 constraints arising from the current experimental limits on the
 electron and on the neutron EDMs are imposed. After the imposition 
 of the constraints the effects of CP violating phases are still 
 quite significant in that the $\chi -p$ cross-section can 
 vary by a factor of $\sim 2$. Thus precision predictions of the
 $\chi -p$ cross-section should take account of the CP phases 
 if indeed such phases do exist in a given model. 
 Indeed many string and  D brane models do indeed possess such
 phases and thus inclusion of such phases is imperative in 
 making predictions for direct detection in such models.
 
\subsection{Effects of coannihilation}
The effects of coannihilation may become important when the 
next to the lowest supersymmetric particle (NLSP) has a mass
which lies close to the LSP mass\cite{yamaguchi0}. 
The size of the effects  is
exponentially damped by the factor $e^{-\Delta_ix}$ where
$\Delta_i=(m_i/m_{\chi}-1)$, $x=m_{\chi}/kT$ and where $m_{\chi}$ 
is the LSP mass. Because of this damping the coannihilation 
effects are typically
important only for regions of the parameter space where
the constraint $\Delta_i< 0.1$ is satisfied. 
 Some of the possible
candidates for NLSP are the light stau $\tilde \tau_1$, 
$\tilde e_R$, the next to the lightest neutralino $\chi^0_2$,
and the light chargino  $\chi_1^+$. 
An interesting result one finds is that in mSUGRA the
upper limit on the neutralino mass consistent with the current
experimental constraints on the relic density is extended from
200 GeV to 600 GeV\cite{coanni} when the effects of 
$\chi -\tilde \tau$ coannihilation are included.
In Secs.(2.6) and (2.7) we will show that the allowed range of the
neutralino mass can also be extended by inclusion of 
nonuniversalities in the gaugino masses. 

\subsection{Nonuniversality of scalar masses}
The minimal SUGRA model is based on the universality at the 
GUT scale. This includes the universality of the scalar 
masses, of the gaugino masses and of the trilinear couplings
at the GUT scale. In supergravity unified models the universality 
of the soft SUSY breaking parameters arises from the assumption 
of a flat Kahler potential. However, the nature of physics 
at the GUT/Planck scale is not fully understood and a more
general analysis of the soft SUSY breaking sector requires that
one work with a curved Kahler potential\cite{soni,nonuni}. 
Such an analysis 
in general leads to nonuniversalities in the scalar sector
of the theory. However, the nonuniversalities in the scalar
sector cannot be completely arbitrary 
 as there are very stringent constraints
on the system from the limits on the flavor changing neutral 
currents (FCNC). A satisfaction of the constraints requires 
essentially a degeneracy in the scalar masses in the first 
two generations at the GUT scale. However, the constraints on
the scalar masses in the Higgs sector and on masses in the 
third generation 
are far less severe and one could  introduce  significant amounts
of nonuniversalities in these sectors without violating 
the FCNC constraints. It is found convenient to parametrize the 
nonuniversalities in the Higgs sector by  $\delta_1,\delta_2$ so that
 at the GUT scale ($M_G$) one has $m_{H_1}^2 =  m_0^2 (1 + \delta_1)$,  ~~
$m_{H_2}^2 = m_0^2 (1 + \delta_2)$.  Similarly one may parametrize 
the nonuniversalities in the third generation squark sector by $\delta_3,
\delta_4$ so that at the scale
 $M_G$ one has $m_{\tilde Q_L}^2=m_0^2(1+\delta_3)$, ~~
 $m_{\tilde U_R}^2=m_0^2(1+\delta_4)$. These nonuniversalities
 have a significant effect on the low energy physics.
 One of the main effects that occurs is through the effect on $\mu$
 which is determined via the constraint of the radiative breaking of
 the electro-weak symmetry and is modified in the presence of the
 nonuniversalities in the Higgs sector and in the third generation
 sector. To one loop order it is given by\cite{nonuni}
\begin{equation}
\mu^2=\mu_0^2+\frac{m_0^2}{t^2-1}(\delta_1-\delta_2t^2-
\frac{ D_0-1}{2}(\delta_2 +\delta_3+\delta_4)t^2) +\Delta \mu^2
\end{equation}
Here $\mu_0$ is the value of $\mu$ in the absence of nonuniversalities,
$D_0$ depends on the top Yukawa coupling and defines the position 
of the Landau pole, $t\equiv tan\beta$, and $\Delta\mu^2$ is the
loop correction. We note that the entire effect of 
nonuniversalities is now explicity exhibited. 
One finds that the universalities can significantly affect 
the event rates. The effect on the event rates occurs specifically
because of the effect on $\mu$. Thus one finds that for certain
regions of the parameter space the nonuniversalities in the 
Higgs and in the third generation sector make a negative contribution
to  $\mu^2$ which leads to larger higgsino components for the neutralino.
Since in the direct detection the scattering is dominated by 
the  scalar $\chi -p$ cross-section which in turn depends on 
the  product of the gaugino and the higgsino components one  finds that
a smaller $\mu$ leads to larger event rates in the direct  detection.
A detailed analysis of the effects of 
nonuniversalities of the scalar masses has been given in 
Refs.\cite{nonuni}. 
We will discuss further this phenomena in the context of the
 nonuniversalities in the gaugino sector in the next section.
 \subsection{Gaugino nonuniversalities and dark matter in GUT models}
\begin{figure}[t]
 \vspace{8.0cm}
\includegraphics{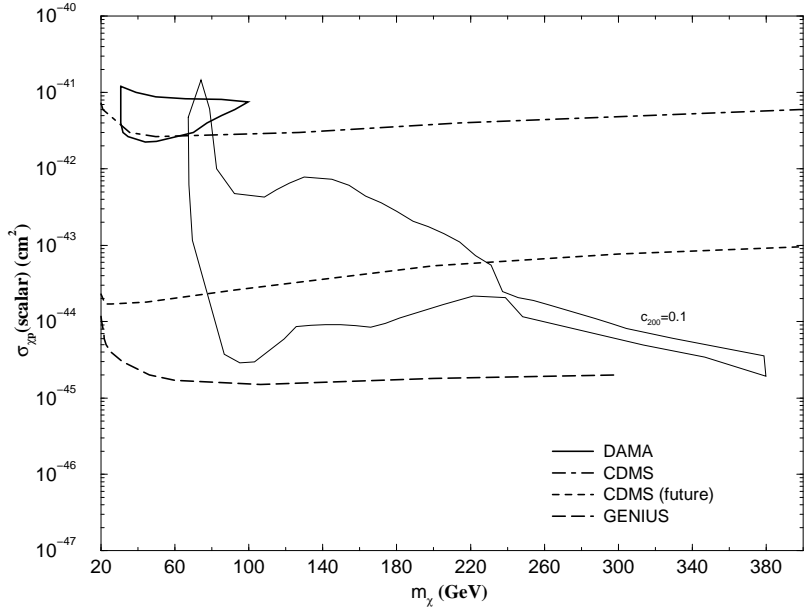}
 \caption{\it
	The maximum and the minimum curves of $\sigma_{\chi p}(scalar)$ vs 
	$m_{\chi}$  when $c_{200}=0.1$ and SUSY parameters are varied
	over the natualness domain (Taken from Ref.\cite{corsetti2}).
\label{exfig2} }
\end{figure}

Nonuniversality of gaugino masses arises in grand unified models 
via corrections to
the gauge kinetic energy functions\cite{hill}.
Thus in grand unified models a non-trivial 
gauge kinetic energy function leads to a gaugino mass matrix
which has the form\cite{hill}
\begin{equation}
m_{\alpha\beta}= \frac{1}{4}\bar{e}^{G/2}G^a(G^{-1})^b_a
(\partial f^*_{\alpha\gamma}
/\partial z^{*b})f^{-1}_{\gamma\beta}
\end{equation}
As  an example if we consider the GUT group to be SU(5) then 
the gauge kinetic energy function $f_{\alpha\beta}$ transforms as
follows
\begin{equation}
({\bf 24}\times {\bf 24})_{symm}={\bf 1}+{\bf 24}+{\bf 75}
+{\bf 200}
\end{equation}
where $({\bf 24}\times {\bf 24})_{symm}$ stands for the symmetric
product. The term that transforms like the singlet of SU(5) in the
gauge kinetic energy function leads to universality of the gaugino masses,
while the ${\bf 24}$ plet, the ${\bf 75}$ plet and the ${\bf 120}$
plet will generate corrections to universality. In general one could
have an admixture of the various representations and this will lead to
gaugino masses of the form 
\begin{equation}
\tilde m_i(0)=m_{\frac{1}{2}}(1+ \sum_r c_r n_i^r)
\end{equation}
where $n_i^r$ depend on r and for the representations ${\bf 1, 24, 75, 210}$
they are given in Table 1\cite{anderson1}. 
The nonuniversality of the gaugino masses
also leads to corrections of the gauge coupling constants at the GUT scale 
and in general one has $g_i(M_G)=g_G (1+ \sum_r c_{r}' n_i^r)$.
We note, however, that the coefficients $c_r'$ that enter in  
$g_i$ are different  than those that enter in $m_i$.
This is so because the corrections to $g_i$ involve only the gauge
kinetic energy function while the corrections to $m_i$ involve 
the gauge kinetic energy function as well as the nature of GUT
physics.
\begin{center} 
\begin{tabular}{|c|c|c|c|}
\multicolumn{4}{c}{Table: nonuniversalities at $M_X$.  } \\
\hline
  SU(5) rep & $n_1^r$ & $n_2^r$& $n_3^r$ \\
\hline
 {\bf 1} & 1 &1&1 \\
 \hline
 {\bf 24} & -1 &$-3$&$2$\\
 \hline
 {\bf 75} &-5&3 &$1$\\
 \hline
 {\bf 200} & 10 &2& 1\\
 \hline
\end{tabular}\\
\noindent
\end{center}
 The gaugino sector nonuniversalities affect $\mu$. To exhibit
 this  effect we can expand $\mu$ determined via the constraint
 of the radiative breaking of the electro-weak symmetry in terms
 of the parameter $c_r$. One finds the following expansion
  \begin{equation}
  \tilde \mu^2= \mu^2_{0}+\sum_r \frac{\partial\tilde\mu^2}
  {\partial c_r} c_r+ O(c_r^2)
  \end{equation}
  and for $c_{24}<0, c_{75}<0, c_{200}>0$ 
  one has\cite{corsetti2} 
  \begin{equation}
    \frac{\partial \mu^2_{24}}{\partial c_{24}} >0,
  \frac{\partial \mu^2_{75}}{\partial c_{75}} >0,
  \frac{\partial \mu^2_{200}}{\partial c_{200}}<0 
  \end{equation} 
 Thus in these cases the  nonuniversalities lead to a 
  smaller value of $|\mu|$. Now as already mentioned in the
  previous section the Higgsino components become more dominant
  as $\mu$ becomes smaller. We can exhibit this analytically for
  the case when $\mu$ is small but we are still in the 
  scaling region\cite{scaling1}
  where $\mu^2/M_Z^2>>1$. In this  case it is possible to
  analytically investigate the size of the gaugino-Higgsino 
  components $X_{n0}$ of the LSP defined by  
 \begin{equation}
\chi=X_{10} \tilde B+ X_{20}\tilde W_3 + X_{30}\tilde H_1
+ X_{40} \tilde H_2
\end{equation}
  where $\tilde B$ is the Bino, $\tilde W_3$ is the Wino, and
  $\tilde H_1$, and $\tilde H_2$ are the two Higgsinos.
 In this case one finds that the
  gaugino components of the LSP are given by 
  $X_{11}\simeq 1-(\frac{M_Z^2}{2\mu^2})sin^2\theta_W$,
 and $X_{12}\simeq\frac{M_Z^2}{2m_{\chi_1}^2\mu}sin2\theta_W sin\beta$
 while the higgsino components are given by\cite{scaling1}
$X_{13}\simeq -\frac{M_Z}{\mu}sin2\theta_W sin\beta$,
$X_{14}\simeq \frac{M_Z}{\mu}sin2\theta_W sin\beta$.
From the above one finds that the Higgsino components have a
dependence on the inverse power of $\mu$ and thus a smaller $\mu$
will lead to a larger scalar $\sigma_{\chi -p}$ cross-section.
\begin{figure}[t]
 \vspace{8.0cm}
\includegraphics{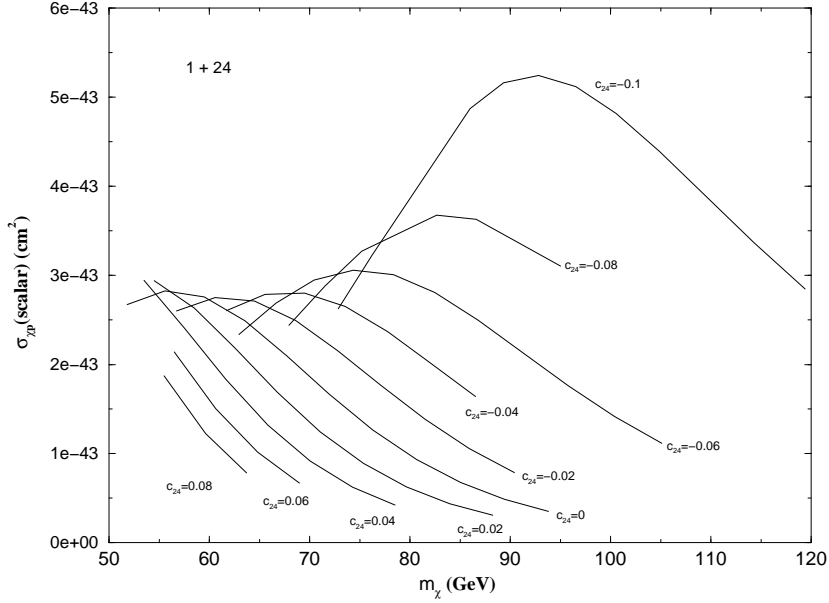}
 \caption{\it
       $\sigma_{\chi p}(scalar)$ vs $m_{\chi}$  when $m_0=51$ GeV,
       $\tan\beta=10$, $A_t/m_0=-7$  and $c_{24}$ takes on various
       values (Taken from Ref.\cite{corsetti2}).
    \label{exfig1} }
\end{figure}
The literature on the analyses of dark matter  
relic density\cite{goldberg,accurate} and 
direct detection in MSSM\cite{witten} and in SUGRA models\cite{sugra}
is quite extensive\cite{jungman}.  
We discuss here the quantitative effects of the gaugino mass
nonuniversality on dark matter. Some features of the effects of
gaugino mass nonuniversalities have already been discussed in
the literature\cite{greist} and we review here the more recent 
developments\cite{corsetti2}.
The techniques used
in the analysis are as discussed in Ref.\cite{sugra}
and in the analysis we impose the $b\rightarrow s+\gamma$ 
constraint\cite{bsgamma}. In Fig.\ref{exfig1} we plot  the 
scalar $\chi -p$ cross-section as  a function of the neutralino
mass for the case of GUT scale nonuniversalities with 
values of $c_{24}$ in the range  -0.1 to 0.08. One finds 
that the scalar cross-section is enhanced for negative values of
$c_{24}$ just as one would expect from the general discussion
above  because it is for the case of $c_{24}$ negative that 
$\mu$ becomes small. One  finds that in general the scalar
cross-section increases systematically  as  $|c_{24}|$
increases for negative values of $c_{24}$ and an enhancement of
the scalar cross-section by as much as a factor of 10 can be gotten
relative to the universal case of $c_{24}=0$. One also finds 
an enhancement of the allowed range of the neutralino mass  consistent
with the constraints.  In Fig.\ref{exfig2} we plot the maximum and the minimum 
of the scalar cross-section as a function of the neutralino mass
for the case of GUT scale nonuniversalities where the nonuniversalities
arise from the 200 plet representation with $c_{200}=0.1$
when the other parameters  are varied over their assumed naturalness
range.  The current experimental limits from DAMA\cite{dama} and 
from CDMS\cite{cdms} are also plotted. Further, the currents limits
would certainly be significantly improved in other dark matter detectors
in the future\cite{cdms,spooner,baudis} and  in Fig.\ref{exfig2} we
also plot the expected limits from future CDMS, and from GENIUS\cite{baudis}. 
One finds that the current experiment does constrain 
the theory in a small region of the parameter space. Further,
the expected sensitivity in future experiment, i.e., in CDMS and
in GENIUS will explore  a major part of the parameter space of this
model. We also note that the inclusion of nonuniversality 
significantly increases the allowed range of the neutralino 
parameter space.
\subsection{ Dark Matter on D Branes} 
    Nonuniversality of gaugino masses is rather  generic in 
    string theory. However, the specific nature of the 
    nonuniversality will depend on the details of the 
    compactification.
    We discuss here the effects of gaugino mass nonuniversality on 
    dark matter in the context of D brane 
    models.  
 The possibility of nonuniversal gaugino phases in brane models arises from 
 the choice of embedding of the different gauge groups of the
 Standard Model on different branes. One may consider, 
 from example, models that arise from Type IIB string compactified
 on a six-torus $T^2\times T^2\times T^2$ which contains 9 branes,
 $7_i$ and $5_i$ (i=1,2,3) branes and 3 branes. 
 Not all the branes can be present simultaneuosly due to
 the constraint of N=1 supersymmetry which requires that one has either 
 9 branes and $5_i$ (i=1,2,3) branes or $7_i$ (i=1,2,3) branes 
 and 3 branes. In the following we will make the choice of embedding
 on $9$ branes  and 5 branes\cite{in2}. 
 One of the major problems in 
 developing a sensible string phenomenology is that  the mechanism
 of supersymmetry breaking in string theory is still lacking. 
 However, some progress can be made  by use of an efficient 
 parametrization of supersymmetry breaking. Here we use the
 parametrization where the breaking of supersymmetry arises from
 the breaking generated by the dilaton and the moduli VEV's of 
 the following form\cite{ibanez}
 $F^S=$$\sqrt 3 m_{3/2}$$ (S+S^*)sin\theta e^{-i\gamma_S}$,~~
 $F^i=$$\sqrt 3 m_{3/2}$$ (T_i+T_i^*)cos\theta_i e^{-i\gamma_i}$.
We consider now a specific 9-5 brane model. Here one embeds 
the $SU(3)_C\times U(1)_Y$ gauge group on  9 branes  and 
$SU(2)_L$ gauge group on a $5_1$ brane. The alternative possibility
of embedding the Standard Model gauge  group on five branes is
discussed in the last two papers of Ref.\cite{in1}.
  For the $9-5_1$ brane model the soft SUSY breaking sector of the theory 
is given by\cite{ibanez,in2} 
\begin{eqnarray}
\tilde m_1=\tilde  m_3=\sqrt 3 m_{3/2}sin\theta e^{-i\gamma_S}=-A_0;~~
 \tilde m_2=\sqrt 3 m_{3/2} cos\theta e^{-i\gamma_i}\nonumber\\
 \tilde m_9^2=m_{3/2}^2(1-3cos^2\theta \Theta_1^2);~~
 \tilde m_{95_1}^2=m_{3/2}^2(1-(3/2)cos^2\theta(1-\Theta_1^2))
 \end{eqnarray}
\begin{figure}[t]
 \vspace{8.0cm}
\includegraphics{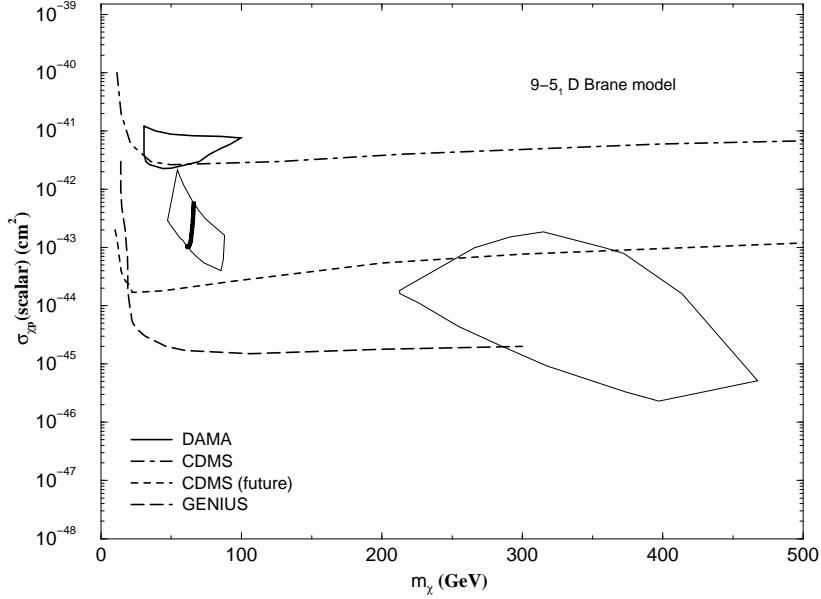}
 \caption{\it
 	The maximum and the minimum curves of $\sigma_{\chi p}(scalar)$ vs 
 	$m_{\chi}$  for the 9-5 D brane model when $c_{200}=0.1$ when 
 	$m_{3/2}$ ranges up to 2 TeV, $\tan\beta$ ranges up to 25, and 
 	$\theta$ lies in the range 0.1-1.6. (Taken from Ref.\cite{corsetti2}).
\label{exfig3} }
\end{figure}
Here $\theta (\Theta_i)$ is the Goldstino direction in the dilaton S
 (moduli $T_i$) VEV space.
   We discuss now dark matter on D branes. In Fig.\ref{exfig3} we give
  a plot of the scalar $\chi -p$ cross section as a function
  of the neutralino mass for the $9-5_1$ D brane model. 
  One of the interesting feature of the D brane  model is 
  that the scalar masses are in general not universal. 
  However for $\Theta_1=1/\sqrt 3$ one has $m_9 = m_{95_1}$
  and the scalar masses are universal although the 
  gaugino masses are still nonuniversal. Since we are mostly
  interested here in investigating the effects of 
  nonuniversalities of the gaugino masses in this analysis, we 
  impose universality of the scalar masses and set 
  $\Theta_1=1/\sqrt 3$. In Fig.\ref{exfig3} we give a plot of the 
  minimum and the maximum of the scalar $\chi -p$ cross section
  under this constraint. One finds that under the assumed constraints  
  the allowed domain of the parameter space has the general 
  features which are similar to the GUT scale nonuniversalities.
  One common feature is that the allowed domain of the parameter
  space is extended close to 500 GeV. One may note that if
  in addition to the constraint $\Theta_1=1/\sqrt 3$ one also 
  sets $\theta =\pi/6$ one finds also universality of the 
  gaugino masses. This  situation is exhibited by the vertical
  dark  line in the enclosed region on the left hand side in Fig.\ref{exfig3}.
\section{Conclusion}
In this paper we have given a brief review of the recent theoretical
developments in the analyses of supersymmeteric dark matter.
Our emphasis has been in exploring the effects of uncertainties 
of the input data and the effects of nonuniversalities 
of the gaugino masses on dark matter analyses. It is found that the
uncertainties of the wimp velocities can change detection rates
by up to factors of 2-3 while the uncertainties in quark masses
and densities can change the $\chi -p$ cross-section by up to factors
of 5 in either direction. The effects of gaugino mass nonuniversalities
on dark matter analyses is found to be quite dramatic. It is seen that 
gaugino mass nonuniversalities can increase the Higgsino components 
of the LSP and significantly increase the $\chi -p$ cross-section 
from scalar interactions and also increase the allowed range of the 
LSP consistent 
with relic density constraints.  Thus an increase in the scalar  
 $\chi -p$ cross-sections by up to a factor of 10 can occur while
 the allowed range of the neutralino masses can move up to 500 GeV
 consistent with the relic  density constraints.
 Data from current dark matter experiments is beginning to put 
 constraints on models with nonuniversalities. These constraints 
 will become more severe as the sensitivity of dark matter 
 experiments increase in the future. \\	
 \noindent
 {\bf Acknowledgements}\\ 	   	
  This research was supported in part by NSF grant 
PHY-9901057.

\end{document}